\documentclass[final,5p,times,twocolumn]{elsarticle}

\usepackage{amssymb}
\usepackage{amsmath}
\usepackage{subfigure}
\usepackage{tabularx}
\usepackage{booktabs}
\usepackage{multirow}
\usepackage{threeparttable}
\usepackage[colorlinks, linkcolor=blue, anchorcolor=blue, citecolor=blue]{hyperref}

\journal{arXiv}

\begin{document}

\begin{frontmatter}



\title{EEG-DBNet: A Dual-Branch Network for Temporal-Spectral Decoding in Motor-Imagery Brain-Computer Interfaces\tnoteref{t1}}
\tnotetext[t1]{This work was supported by the National Natural Science Foundation of China [grant number 62171073, 62311530103, 62106032 and 62236002]; “Chunhui Plan" Collaborative Research Project of the Ministry of Education, China [grant number HZKY20220209]; Natural Science Foundation of Chongqing, China [grant number CSTB2023NSCQ-LZX0064]; Chongqing Scientific Research Innovation Project for Postgraduate Students [grant number CYB23241]; and the Doctoral Training Program of Chongqing University of Posts and Telecommunications [grant number BYJS202317].}

\author[label1]{Xicheng Lou}
\author[label2]{Xinwei Li}
\author[label3]{Hongying Meng}
\author[label4]{Jun Hu}
\author[label4]{Meili Xu}
\author[label1]{Yue Zhao}
\author[label5]{Jiazhang Yang}
\author[label2]{Zhangyong Li\corref{cor1}}
\ead{lizy@cqupt.edu.cn}

\affiliation[label1]{organization={School of Communication and Information Engineering, Chongqing University of Posts and Telecommunications},
	city={Chongqing},
	postcode={400065},
	country={China}}

\affiliation[label2]{organization={Research Center of Biomedical Engineering, Chongqing University of Posts and Telecommunications},
	city={Chongqing},
	postcode={400065},
	country={China}}
	
\affiliation[label3]{organization={Department of Electrical, Brunel University London},
	city={London},
	postcode={UB8 3PH},
	country={United Kingdom}}

\affiliation[label4]{organization={Department of Neurology, Southwest Hospital, Army Medical University},
	city={Chongqing},
	postcode={400038},
	country={China}}
	
\affiliation[label5]{organization={Yongchuan Women and Children Hospital},
	city={Chongqing},
	postcode={402160},
	country={China}}

\cortext[cor1]{Corresponding author}

%

\begin{abstract}

Motor imagery electroencephalogram (EEG)-based brain-computer interfaces (BCIs) offer significant advantages for individuals with restricted limb mobility. However, challenges such as low signal-to-noise ratio and limited spatial resolution impede accurate feature extraction from EEG signals, thereby affecting the classification accuracy of different actions. To address these challenges, this study proposes an end-to-end dual-branch network (EEG-DBNet) that decodes the temporal and spectral sequences of EEG signals in parallel through two distinct network branches. Each branch comprises a local convolutional block and a global convolutional block. The local convolutional block transforms the source signal from the temporal-spatial domain to the temporal-spectral domain. By varying the number of filters and convolution kernel sizes, the local convolutional blocks in different branches adjust the length of their respective dimension sequences. Different types of pooling layers are then employed to emphasize the features of various dimension sequences, setting the stage for subsequent global feature extraction. The global convolution block splits and reconstructs the feature of the signal sequence processed by the local convolution block in the same branch, and further extracts features through the dilated causal convolutional neural networks. Finally, the outputs from the two branches are concatenated, and signal classification is completed via a fully connected layer. Our proposed method achieves classification accuracies of $85.84\%$ and $91.60\%$ on the BCI Competition \uppercase\expandafter{\romannumeral4}-2a and BCI Competition \uppercase\expandafter{\romannumeral4}-2b datasets, respectively, surpassing existing state-of-the-art models. The source code is available at \url{https://github.com/xicheng105/EEG-DBNet}.

\end{abstract}




\end{frontmatter}


\section{Introduction}
\label{sec:introduction}



As a direct pathway for information exchange between the human brain and computers, brain-computer interfaces (BCIs) translate neural activity into commands for controlling external devices \citep{lahane2019review}. Typically, BCIs extract features from electroencephalogram (EEG) signals, measured by electrodes placed on the scalp, and convert them into outputs to control devices such as wheelchairs and displays. Motor imagery EEG (MI-EEG) is a spontaneous electrical potential generated when subjects imagine movements without actual movement. MI-EEG-based BCIs have wide applications in motor control \citep{pirondini2017eeg}, neural rehabilitation \citep{orban2022review}, and specialized environmental operations \citep{natheir2023utilizing}.

However, decoding MI-EEG poses several challenges. The limited number of electrodes in acquisition devices results in low spatial resolution. The prevalent use of non-invasive methods, driven by safety and portability, introduces isolation between electrodes and the brain, leading to a low signal-to-noise ratio (SNR). The low spatial resolution makes methods that transform signal sequences into topological maps highly susceptible to the influence of the number of electrodes, with more electrodes generally improving accuracy \citep{zhao2019multi}. Meanwhile, the low SNR increases the likelihood of models mistaking noise for features during training, necessitating robust feature extraction capabilities. Consequently, data preprocessing is crucial for traditional machine learning-based EEG decoding methods, including linear discriminant analysis (LDA) \citep{fraiwan2012automated}, support vector machine (SVM) \citep{subasi2010eeg}, naive Bayesian classifier (NBC) \citep{machado2014executed}, and K-nearest neighbor (KNN) \citep{sha2020knn}. Preprocessing techniques such as filtering \citep{ang2012filter}, transformation \citep{bashar2015motor}, artifact removal \citep{jafarifarmand2019eeg}, and electrode selection \citep{zhu2019study} require specialized knowledge and significant computational resources.

Deep neural networks offer an integrated approach to data preprocessing, providing an end-to-end solution for EEG signal decoding and becoming the mainstream method in the field. Convolutional neural networks (CNNs), known for their feature extraction capabilities in image processing, have been applied to EEG decoding. \citet{schirrmeister2017deep} proposed two models with different complexities: Deep ConvNet and Shallow ConvNet. Inspired by the filter bank common spatial pattern (FBCSP) \citep{ang2008filter}, these models augmented the pseudo-spectral dimension of raw signals, simulated spectral characteristics via channel-wise convolution, and compressed the spatial dimension. \citet{lawhern2018eegnet} enhanced the Shallow ConvNet by introducing depth-wise separable convolution from MobileNet \citep{howard2017mobilenets}, creating EEGNet, a compact and efficient network that has become a benchmark in EEG decoding models, with numerous subsequent models building upon its structure for refinement. \citet{chen2024toward} proposed EEG-NeX, which builds upon EEGNet by increasing the number of convolutional layers to extract more temporal feature maps. They replaced separable convolutions with traditional convolutions to enhance feature extraction accuracy and designed an inverse bottleneck structure for the convolutional channels to achieve feature reconstruction. Additionally, they introduced dilated convolutions \citep{oord2016wavenet} to improve the network's receptive field. EEG-NeX demonstrated a significant improvement in classification accuracy compared to EEGNet. \citet{ingolfsson2020eeg} proposed EEG-TCNet, which constructed dilated causal convolutional neural networks (DCCNNs) using temporal convolutions \citep{lea2017temporal} and dilated convolutions to enhance feature extraction based on EEGNet. Since dilated convolutions enhance the receptive field by increasing network depth, EEG-TCNet incorporated a residual structure \citep{he2016deep} to avoid network degradation issues caused by deeper networks. \citet{altaheri2022physics} introduced ATCNet, which further builds on EEG-TCNet by incorporating a sliding window for data augmentation and leveraging a multi-head attention mechanism \citep{vaswani2017attention} to strengthen the extraction of global signal features. ATCNet significantly improved the classification accuracy of signals in the BCI Competition \uppercase\expandafter{\romannumeral4}-2a dataset.  

Some scholars believe that the scale of EEG signal samples can no longer match the depth of the model's network. They tend to improve model performance by increasing the number of network branches. \citet{salami2022eeg} introduced the Inception structure \citep{szegedy2015going}, utilizing different convolutional kernels to construct three branches based on EEGNet, thereby increasing the diversity of features extracted from EEG signals. \citet{altuwaijri2022multi} further enhanced the model by incorporating squeeze-and-excitation (SE) modules \citep{hu2018squeeze} to reconstruct the relationships between signals within the three-branch network. \citet{tao2023adfcnn} developed two network branches using separable convolutions and traditional convolutions, respectively, to extract different spectral and spatial information. The branching structure enables models to simultaneously construct multiple signal decoding paradigms, thereby improving model performance through multi-dimensional signal decoding. Since the original EEG signals are spatiotemporal two-dimensional signals lacking a spectral dimension, some researchers have constructed the spectral dimension using traditional signal processing techniques instead of convolution. \citet{mane2021fbcnet} increased the dimensionality of the source EEG through bandpass filtering, extracted features via convolution, and aggregated temporal information through a variance layer. \citet{li2021temporal} extracted multi-frequency band information from EEG using wavelet transform and reconstructed this information using SE modules. \citet{zhi2023multi} expanded the spectral domain channels through filtering and subsequently extracted signal features in three dimensions: spatial, frequency, and time-frequency domains. \citet{hsu2023eeg} employed wavelet transform to reconstruct signal features, enhancing the discrimination of time-frequency maps.

Significant progress has been made in the decoding techniques for MI-EEG, but further improvements are necessary. First, the advantage of branch networks lies in their ability to decode multiple dimensions of a signal simultaneously. Simply reconstructing multiple temporal sequences using each branch to enhance the extraction of temporal features does not fully leverage the strengths of branch networks. Moreover, generating spectral sequences using traditional signal processing methods will be re-modeled during the decoding process, which is not conducive to branch networks training filters based on temporal information to automatically generate spectral sequences that are adapted for subsequent processing steps. Second, convolution and pooling processes compress the sequence length, serving as a feature extraction process. After compression, each sample point on the sequence represents an aggregation of adjacent samples of the source signal, encapsulating local features of the original sequence. However, the characteristics of different types of MI-EEG are expressed through the entire sequence, meaning that local features alone cannot adequately represent MI-EEG.

This study proposes a dual-branch network called EEG-DBNet, where two network branches parallelly decode the temporal and spectral sequences of the MI-EEG. Each branch consists of a local convolution (LC) block and a global convolution (GC) block. The LC block re-models the signal sequences of different dimensions, extracting local features and providing source signals for the GC block. The GC block extracts global features specific to different dimension sequences. The main contributions of this study are as follows:

\begin{enumerate}
	\item We propose a novel dual-branch network, EEG-DBNet, which combines parallel multi-dimensional decoding and serial multi-scale feature extraction techniques to improve the classification accuracy of MI-EEG signals.
	\item For the branches handling signal sequences of different dimensions, we arranged different numbers of filters and various sizes of convolution kernels in the LC blocks to reconstruct the signal sequences. Additionally, we optimized the feature representation of signal sequences of different dimensions by employing different types of pooling layers. The LC blocks effectively extract local features of the signals. Compared to the single-branch structure of EEGNet \citep{lawhern2018eegnet}, there is a significant improvement in the accuracy of signal classification.
	\item We enhanced DCCNNs with sequence splitting and feature reconstruction techniques, constructing GC blocks to perform global feature extraction on the sequences reconstructed by LC blocks. Sequence splitting enables DCCNNs to simultaneously model multiple feature extraction paradigms; feature reconstruction for each split subsequence strengthens the expression of subsequence features, thereby improving the efficiency of DCCNNs. Introducing GC blocks after LC blocks significantly enhances model performance.
	\item Our proposed EEG-DBNet achieved classification accuracies of $85.58\%$ and $91.60\%$ in signal classification experiments on the public datasets BCI Competition \uppercase\expandafter{\romannumeral4}-2a \citep{brunner2008bci} and BCI Competition \uppercase\expandafter{\romannumeral4}-2b \citep{leeb2008bci}, respectively. These results surpass previous models and achieve state-of-the-art performance.
\end{enumerate}

The remainder of this paper is organized as follows: \autoref{sec:Method} details the proposed EEG-DBNet model; \autoref{sec: Experiments} discusses the experiments and their results; \autoref{sec: Conclusion and Discussion} provides the conclusions of this study.

\section{Method}
\label{sec:Method}

\begin{figure}[!h]
	\centering
	\includegraphics{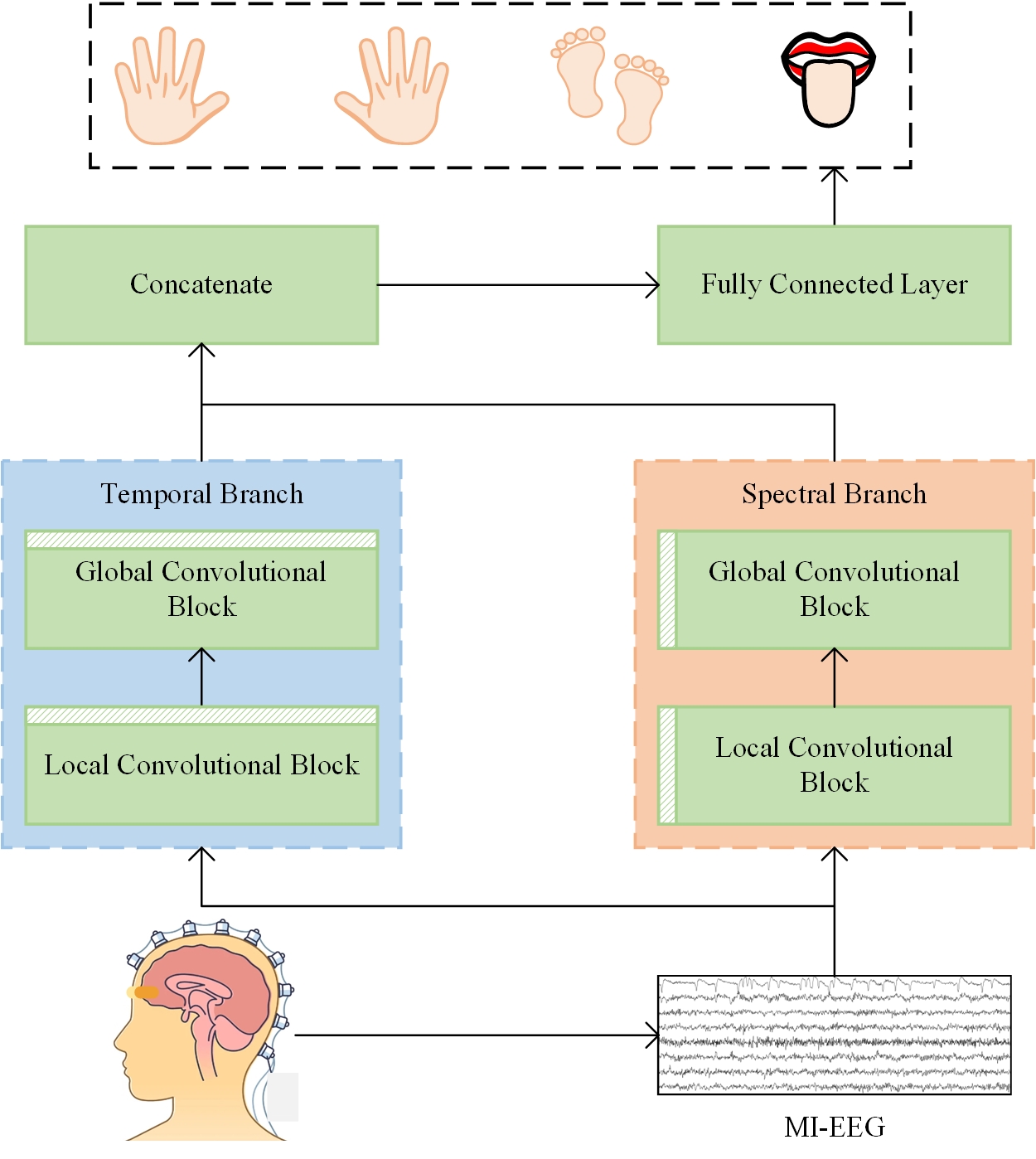}
	\caption{Structure of the proposed EEG-DBNet model.}
	\label{fig_1}
\end{figure}

The structure of EEG-DBNet is depicted in \autoref{fig_1}. The network consists of two branches, each comprising a Local Convolutional (LC) block and a Global Convolutional (GC) block. After the signal is processed through these branches, the outputs are concatenated and classified using a fully connected (FC) layer. 

The input signal $\mathbf{X}_i \in \mathbb{R}^{C \times T}$ for EEG-DBNet is the standardized original signal $\mathbf{Z}_i \in \mathbb{R}^{C \times T}$, where trial $i \in \left\lbrace 1, 2, \cdots, m \right\rbrace$, $T$ denotes the number of sample points, $C$ is the number of EEG electrodes, and $m$ is the number of trials
\begin{align}
	\label{eq1}
	\mathbf{x}^j_{\text{train},i}&=\frac{\mathbf{z}^j_{\text{train},i}-\mu^j_{\text{train},i}}{\sigma^j_{\text{train},i}}\\
	\label{eq2}
	\mathbf{x}^j_{\text{test},i}&=\frac{\mathbf{z}^j_{\text{test},i}-\mu^j_{\text{train},i}}{\sigma^j_{\text{train},i}}
\end{align}
Here, $\mathbf{z}^j_{\text{train},i} \in \mathbb{R}^{1 \times T}$ and $\mathbf{z}^j_{\text{test},i} \in \mathbb{R}^{1 \times T}$ are the training and testing samples for electrode $j \in \left\lbrace 1, 2, \cdots, C \right\rbrace$. $\mathbf{x}^j_{\text{train},i} \in \mathbb{R}^{1 \times T}$ and $\mathbf{x}^j_{\text{test},i} \in \mathbb{R}^{1 \times T}$ are the corresponding standardized samples. $\mu^j_{\text{train},i}$ and $\sigma^j_{\text{train},i}$ represent the mean and variance of the training samples, respectively. EEG-DBNet is trained on the set $\left\lbrace \mathbf{X}_{\text{train},i}, y_i \right\rbrace_{i=1}^m$, where $y_i \in \left\lbrace 1, 2, \cdots, n \right\rbrace$ is the label of $\mathbf{X}_i$. The trained model is then used for classification on $\mathbf{X}_{\text{test},i}$.

\subsection{Local Convolutional Block}
\label{sec: Local Convolutional Block}

The structure of the LC block in each branch is similar to that of EEGNet \cite{lawhern2018eegnet}, as shown in \autoref{fig_2}. It consists of three convolutional layers: a basic convolutional layer, a depthwise convolutional layer, and a separable convolutional layer, along with two pooling layers. Batch normalization (BN) \citep{ioffe2015batch} is applied after the first convolutional layer to accelerate convergence. BN and the exponential linear unit (ELU) \citep{clevert2015fast} are applied after the second and third convolutional layers. The dropout rate \citep{garbin2020dropout} is set to 0.3.

Figure \autoref{fig_2a} illustrates the LC block structure in the temporal branch. The first convolutional layer applies temporal convolution using $\hat{F}_1$ filters of size $(1, \hat{K})$, where $\hat{F}_1 = 8$ and $\hat{K}$ is set to one-fifth of the sampling frequency, which is $250\text{Hz}$ in BCI Competition \uppercase\expandafter{\romannumeral4}-2a and BCI Competition \uppercase\expandafter{\romannumeral4}-2b. Thus, $\hat{K} = 48$. This temporal convolution extracts features above $5\text{Hz}$, producing $\hat{F}_1$ temporal feature maps.

The second convolutional layer performs depthwise convolution with depth $D = 2$ and $\hat{F}_1 \times D$ filters of size $(C, 1)$. This depthwise convolution extracts $D$ spatial features from each temporal feature map, generating a total of $\hat{F}_1 \times D$ feature maps. Next, an average pooling layer of size $(1, \hat{K}/8)$ further integrates the temporal features, reducing the signal's sampling frequency to approximately $42\text{Hz}$ ($250/\hat{K}/8$).

The third convolutional layer employs $\hat{F}$ filters of size $(1, \hat{K}/4)$ to perform separable convolution, where $\hat{F} = \hat{F}_1 \times D$. This separable convolution models the interrelationships of feature maps with a minimal number of parameters and compresses the temporal feature extraction range to approximately $286\text{ms}$ ($1000 \times \hat{K}/4/42$). Finally, another average pooling layer of size $(1, \hat{K}/8)$ reduces the sampling frequency to $7\text{Hz}$ ($42/\hat{K}/8$).

Each $\hat{F}$ corresponds to a temporal sequence at a specific frequency, so sample points at the same position in the time-domain sequences for all $\hat{F}$ form a spectral sequence. The input signal $\mathbf{X}_i \in \mathbb{R}^{C \times T}$ is processed by the temporal branch LC block, resulting in the output $\hat{\mathbf{X}}_{L, i} \in \mathbb{R}^{\hat{F} \times \hat{T}}$. The length of $\hat{T}$ can be determined as:
\begin{figure*}[!th]
	\centering
	\subfigure[]{\includegraphics{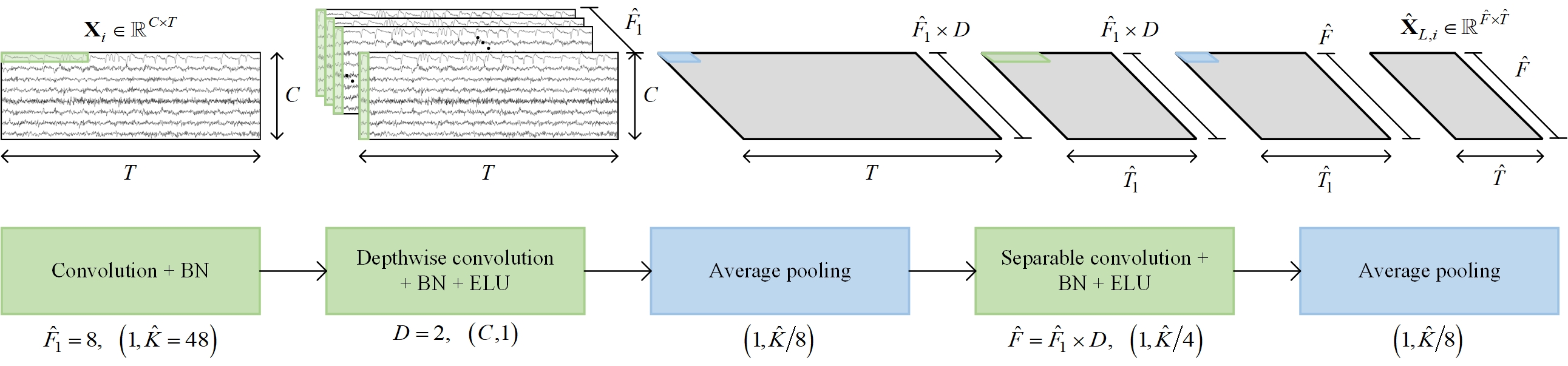}\label{fig_2a}}\\
	\subfigure[]{\includegraphics{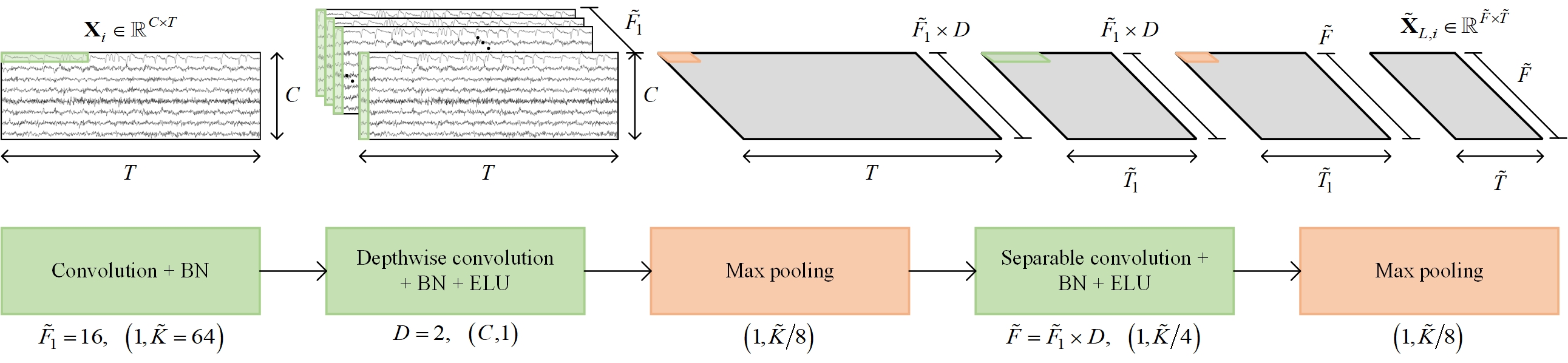}\label{fig_2b}}
	\caption{Structure of the local convolutional blocks. \subref{fig_2a} local convolutional block in the temporal branch. \subref{fig_2b} local convolutional block in the spectral branch.}
	\label{fig_2}
\end{figure*}

\begin{equation}
	\label{eq3}
	\hat{T}=\left\lfloor \frac{64\times T}{\hat{K}^2} \right\rfloor
\end{equation}

The structure of the spectral branch LC block, as shown in Figure \autoref{fig_2b}, includes twice the number of filters compared to the temporal branch, specifically $\tilde{F}_1=16$, to enhance the feature representation of the spectral sequence. Consequently, the length of the spectral sequence also doubles to $\tilde{F}=32$. The filter size is increased to a quarter of the sampling frequency, i.e., $\tilde{K}=64$, which narrows the time-domain range for feature extraction and thereby shortens the length of the temporal sequence $\tilde{T}$. The length $\tilde{T}$ can be determined as follows:
\begin{equation}
	\label{eq4}
	\tilde{T}=\left\lfloor \frac{64\times T}{\tilde{K}^2} \right\rfloor
\end{equation} 

The output of this branch can be denoted as $\tilde{\mathbf{X}}_{L, i}\in\mathbb{R}^{\tilde{F}\times\tilde{T}}$. This adjustment results in an increased spectral sequence length and a decreased temporal sequence length from the spectral branch, aligning its dimensions closely with those of the output signal from the temporal branch, thereby maintaining the symmetry of the branch network. Moreover, as demonstrated by the experiments in section \ref{sec: Ablation Experiments}, for the spectral sequence, the importance of key sample points in expressing sequence features outweighs the relationships between sequence sample points. Therefore, the spectral branch employs max-pooling layers.

\subsection{Global Convolutional Block}
\label{sec: Global Convolutional Block}

The GC block extracts global features from the output of the LC block, as shown in \autoref{fig_3}. A sliding window (SW) with a stride of 1 first segments the signal sequences from the corresponding network branch into $n$ subsequences. The temporal branch signal sequence is denoted as $\hat{T}$, and the length of each subsequence is given by:
\begin{equation}
	\label{eq5}
	\hat{l}=\hat{T}-n+1
\end{equation}
Thus, The input $\hat{\mathbf{X}}_{L, i}\in\mathbb{R}^{\hat{F}\times\hat{T}}$ of the GC block in the temporal branch is split into $\hat{\mathbf{X}}_{L, N, i}\in\mathbb{R}^{\hat{F}\times(\hat{T}-n+1)},N\in\{1,2,\dots,n\}$, as illustrated in Figure  \autoref{fig_3a}. Feature reconstruction for each subsequence provides diverse samples for subsequent feature mining networks, thereby enriching the mined features. We employ a dimension-reduction version of SE \citep{hu2018squeeze} to reconstruct subsequence features. Temporal subsequences first undergo global average pooling to mask distribution information in the spectral dimension:
\begin{align}
	\label{eq6}
	\begin{aligned}
		\hat{\mathbf{x}}_{\text{av},N,i} & =\hat{f}_\text{av}\left( \hat{\mathbf{X}}_{L,N,i}\right) \\
		& =\frac{1}{\hat{F}}\sum_{j=1}^{\hat{F}}\hat{\mathbf{X}}_{L,N,i}\left( j\right)
	\end{aligned}
\end{align}
Subsequently, two FC layers are employed to reconstruct features. The first FC layer compresses the sequence length to $\lceil ( \hat{T}-n+1) /16 \rceil$, and the second FC layer restores the sequence length to its original state. A rectified linear unit (ReLU) \citep{agarap2018deep} activation function follows the first FC layer to speed up network convergence, while a sigmoid activation function follows the second FC layer to transform the output to a range between 0 and 1, serving as the weight sequence for sampling points. Finally, the sequence is recalibrated through the Hadamard product:
\begin{figure*}[!th]
	\centering
	\subfigure[]{\includegraphics{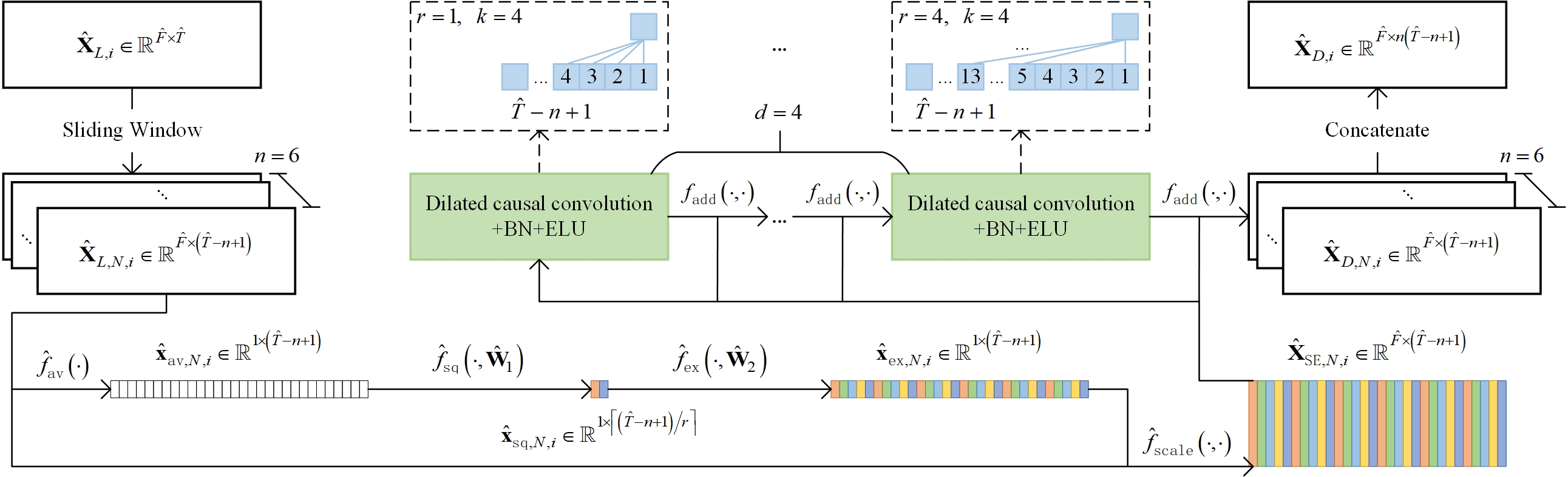}\label{fig_3a}}\\
	\subfigure[]{\includegraphics{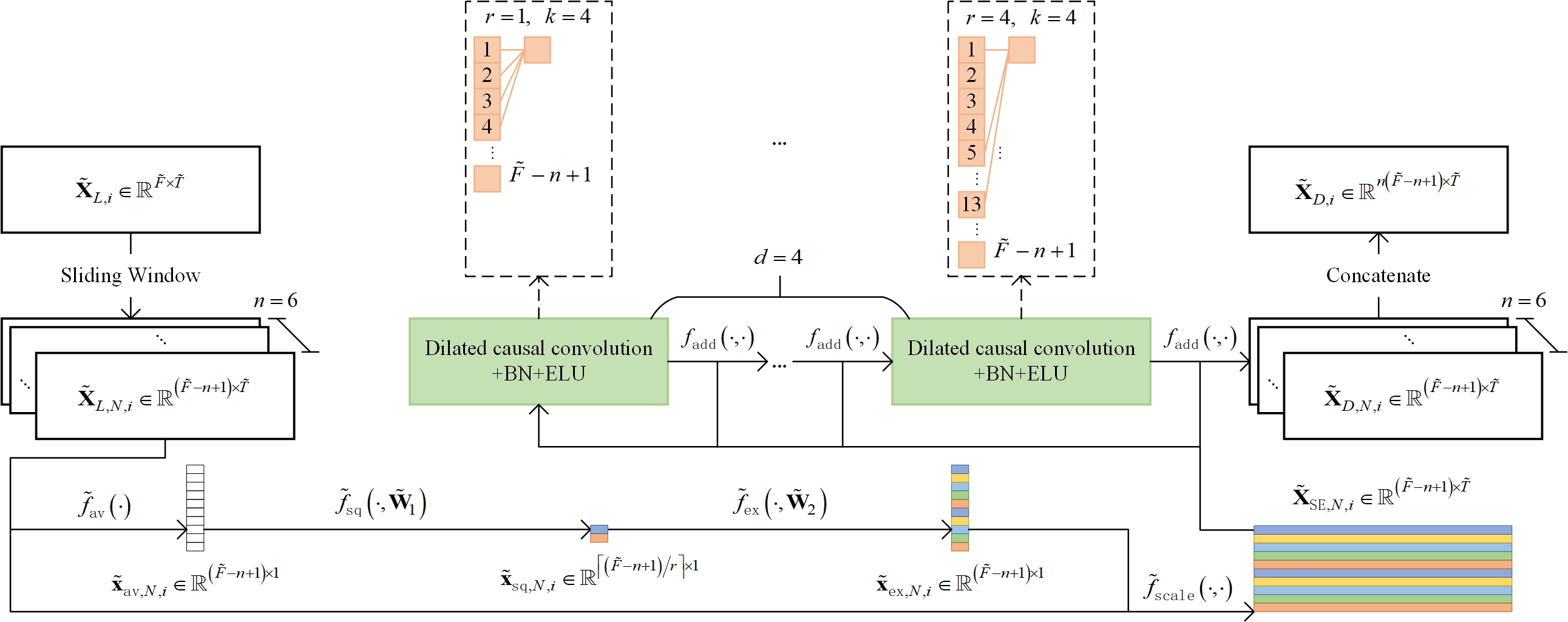}\label{fig_3b}}
	\caption{The structure of the global convolutional blocks with six sliding windows, four dilated causal convolution layers, and the kernel size is set to 4. \subref{fig_3a} global convolutional block in the temporal branch. \subref{fig_3b} global convolutional block in the spectral branch.}
	\label{fig_3}
\end{figure*}

\begin{align}
	\label{eq7}
	\begin{aligned}
		\hat{\mathbf{X}}_{\text{SE},N,i} &=\hat{f}_{\text{scale}}\left( \hat{\mathbf{X}}_{L,N,i}, \hat{f}_{\text{ex}}\left( \hat{f}_{\text{sq}}\left( \hat{\mathbf{x}}_{\text{av},N,i}, \hat{\mathbf{W}}_1 \right), \hat{\mathbf{W}}_2\right) \right) \\
		&=\hat{\mathbf{X}}_{L,N,i}\odot\sigma\left( \hat{\mathbf{W}}_2\delta\left( \hat{\mathbf{W}}_1\hat{\mathbf{x}}_{\text{av},N,i} \right) \right) 
	\end{aligned}		
\end{align}
where $\hat{\mathbf{X}}_{\text{SE},N,i}$ represent the signals after feature reconstruction, $\odot$ denotes the Hadamard product, $\sigma$ represents the sigmoid activation function, $\delta$ denotes the ReLU activation function, and $\hat{\mathbf{W}}_1$, $\hat{\mathbf{W}}_2$ are parameters of the fully connected layers.

The subsequence $\hat{\mathbf{X}}_{\text{SE},N,i}$ after feature reconstruction undergoes $d$ dilated causal convolutional (DCC) layers, each followed by BN and ELU to extract global features. Let the dilation rate $r$ for each DCC layer be $j$, indicating the $j$-th DCC layer, and $k$ be the kernel size. The receptive field $R$ of the DCCNNs can be determined as follows:
\begin{equation}
	\label{eq8}
	R_j=R_{j-1}+\left[ \left( d-1\right) \times\left( k-1\right) +k\right] -1
\end{equation} 
where $R_0=1$. To avoid information loss, the receptive field must be larger than the subsequence length, implying that the hyperparameters of the DCCNNs must satisfy:
\begin{align}
	\label{eq9}
	\begin{aligned}
		R_j & =R_{j-1}+\left[ \left( d-1\right) \times\left( k-1\right) +k\right] -1\\
		& \ge \hat{T}-n+1
	\end{aligned}		
\end{align}
Through the experiments in section \ref{sec: Ablation Experiments}, we determined that the optimal values for $d$ and $k$ in both branches are 4 and 4, respectively. 

The output of each DCC layer is added to the source subsequence before being fed into the next DCC layer:
\begin{align}
	\label{eq10}
	\begin{aligned}
		\hat{\mathbf{X}}_{D,N,i} &= f_\text{add}\left( \hat{\mathbf{X}}_{\text{SE},N,i}, \hat{\mathbf{X}}_{\text{DCC},j}\right) \\
		&=  \zeta\left( \hat{\mathbf{X}}_{\text{SE},N,i}+ \hat{\mathbf{X}}_{\text{DCC},j} \right) 
	\end{aligned}
\end{align}
\begin{align}
	\label{eq11}
	\begin{aligned}
		\hat{\mathbf{X}}_{\text{DCC},j} &= \hat{f}_{\text{DCC},j}\left( f_\text{add}\left( \hat{\mathbf{X}}_{\text{SE},N,i}, \hat{\mathbf{X}}_{\text{DCC},j-1}\right)  \right) \\
		&=  \hat{f}_{\text{DCC},j}\left( \zeta\left( \hat{\mathbf{X}}_{\text{SE},N,i}+ \hat{\mathbf{X}}_{\text{DCC},j-1} \right) \right)
	\end{aligned}
\end{align}
where $j\in\left\lbrace 2,3,\dots,d\right\rbrace $, $\zeta$ refers to the ELU function, $\hat{f}_{\text{DCC},j}$ denotes the $j$-th DCC layer in the temporal branch, and $\hat{\mathbf{X}}_{\text{DCC},j}$ is the out put of this DCC layer. This residual structural is necessary because deep DCCNNs disrupt the sequence, excessively emphasizing backend results, which may strengthen the ``patched'' features. The output $\hat{\mathbf{X}}_{D, i}\in\mathbb{R}^{\hat{F}\times n(\hat{T}-n+1)}$ of the GC block in the temporal branch is obtained by concatenating $n$ processed subsequences $\hat{\mathbf{X}}_{D, N, i}\in\mathbb{R}^{\hat{F}\times(\hat{T}-n+1)},N\in\{1,2,\dots,n\}$.

For the spectral branch, the signal sequence is denoted as $\tilde{F}$. After partitioning using a sliding window, the subsequences can be represented as
\begin{equation}
	\label{eq12}
	\tilde{l}=\tilde{F}-n+1
\end{equation}
The input $\tilde{\mathbf{X}}_{L, i}\in\mathbb{R}^{\tilde{F}\times\tilde{T}}$ of the GC block in the spectral branch is thus split into $\tilde{\mathbf{X}}_{L, N, i}\in\mathbb{R}^{\left( \tilde{F}-n+1\right) \times\tilde{T}},N\in\{1,2,\dots,n\}$, as illustrated in Figure \ref{fig_3b}. The spectral subsequence masks the information in the temporal dimension through global average pooling
\begin{align}
	\label{eq13}
	\begin{aligned}
		\tilde{\mathbf{x}}_{\text{av},N,i} & =\tilde{f}_\text{av}\left( \tilde{\mathbf{X}}_{L,N,i}\right) \\
		& =\frac{1}{\tilde{T}}\sum_{j=1}^{\tilde{T}}\tilde{\mathbf{X}}_{L,N,i}\left( j\right).
	\end{aligned}
\end{align}
The feature reconstruction of the spectral subsequence is achieved by 
\begin{align}
	\label{eq14}
	\begin{aligned}
		\tilde{\mathbf{X}}_{\text{SE},N,i} &=\tilde{f}_{\text{scale}}\left( \tilde{\mathbf{X}}_{L,N,i}, \tilde{f}_{\text{ex}}\left( \tilde{f}_{\text{sq}}\left( \tilde{\mathbf{x}}_{\text{av},N,i}, \tilde{\mathbf{W}}_1 \right), \tilde{\mathbf{W}}_2\right) \right) \\
		&=\tilde{\mathbf{X}}_{L,N,i}\odot\sigma\left( \tilde{\mathbf{W}}_2\delta\left( \tilde{\mathbf{W}}_1\tilde{\mathbf{x}}_{\text{av},N,i} \right) \right) 
	\end{aligned}		
\end{align}
where $\tilde{\mathbf{X}}_{\text{SE},N,i}$ represent the signals after feature reconstruction, and $\tilde{\mathbf{W}}_1$, $\tilde{\mathbf{W}}_2$ are parameters of the fully connected layers. The structure and hyperparameters of the DCCNNs in the spectral branch are identical to those in the temporal branch; the difference is that it processes spectral sequences. Therefore, the receptive field must satisfy
\begin{align}
	\label{eq15}
	\begin{aligned}
		R_j & =R_{j-1}+\left[ \left( d-1\right) \times\left( k-1\right) +k\right] -1\\
		& \ge \tilde{F}-n+1.
	\end{aligned}		
\end{align}
Combining with (\ref{eq9})
\begin{equation}
	\label{eq16}
	R_j\ge\max\left\lbrace \hat{T}-n+1,\tilde{F}-n+1\right\rbrace 
\end{equation}
The output $\tilde{\mathbf{X}}_{D, i}\in\mathbb{R}^{n(\tilde{F}-n+1)\times\tilde{T}}$ of the GC block in the spectral branch is obtained by concatenating $n$ processed subsequences $\tilde{\mathbf{X}}_{D,N,i}\in\mathbb{R}^{(\tilde{F}-n+1)\times\tilde{T}},N\in\{1,2,\dots,n\}$, where
\begin{align}
	\label{eq17}
	\begin{aligned}
		\tilde{\mathbf{X}}_{D,N,i} &= f_\text{add}\left( \tilde{\mathbf{X}}_{\text{SE},N,i}, \tilde{\mathbf{X}}_{\text{DCC},j}\right) \\
		&=  \zeta\left( \tilde{\mathbf{X}}_{\text{SE},N,i}+ \tilde{\mathbf{X}}_{\text{DCC},j} \right) 
	\end{aligned}
\end{align}
\begin{align}
	\label{eq18}
	\begin{aligned}
		\tilde{\mathbf{X}}_{\text{DCC},j} &= \tilde{f}_{\text{DCC},j}\left( f_\text{add}\left( \tilde{\mathbf{X}}_{\text{SE},N,i}, \tilde{\mathbf{X}}_{\text{DCC},j-1}\right)  \right) \\
		&=  \tilde{f}_{\text{DCC},j}\left( \zeta\left( \tilde{\mathbf{X}}_{\text{SE},N,i}+ \tilde{\mathbf{X}}_{\text{DCC},j-1} \right) \right)
	\end{aligned}
\end{align}
$\tilde{f}_{\text{DCC},j}$ denotes the $j$-th DCC layer in the spectral branch.

Vectorize $\hat{\mathbf{X}}_{D, i}$ and $\tilde{\mathbf{X}}_{D, i}$ to obtain $\hat{\mathbf{x}}_i\in\mathbb{R}^{\hat{F}n(\hat{T}-n+1)}$ and $\tilde{\mathbf{x}}_i\in\mathbb{R}^{n(\tilde{F}-n+1)\tilde{T}}$. Concatenate $\hat{\mathbf{x}}_i$ and $\tilde{\mathbf{x}}_i$, and perform signal classification through an FC layer to obtain the predicted label $\check{y}_i$
\begin{equation}
	\label{eq21}
	\check{y}_i = \sigma\left(\mathbf{W}\left(\hat{\mathbf{x}}_i^\frown\tilde{\mathbf{x}}_i\right)\right)
\end{equation}
where $^\frown$ denotes concatenate, $\mathbf{W}$ is parameters of the FC layer.

\section{Experiments}
\label{sec: Experiments}

\subsection{Datasets}
\label{sec: Datasets}

The effectiveness of the proposed EEG-DBNet was validated through extensive experiments on two public datasets: BCI Competition \uppercase\expandafter{\romannumeral4}-2a \cite{brunner2008bci} and \uppercase\expandafter{\romannumeral4}-2b \cite{leeb2008bci}.

BCI Competition \uppercase\expandafter{\romannumeral4}-2a consists of two sessions, each comprising EEG data recorded using 22 electrodes collected from 9 subjects. Each subject participated in a 4-class motor imagery EEG experiment, including left hand (L), right hand (R), feet (F), and tongue (T) imagery. Each subject completed 72 trials for each class in each session, resulting in a total of 576 trials per subject. The time segment for each trial is set to $\left[ 1.5,6\right] $ seconds. In this study, data from the first session serve as the training set, while data from the second session serve as the test set.

BCI Competition \uppercase\expandafter{\romannumeral4}-2b consists of five sessions, each comprising EEG data recorded using three electrodes collected from 9 subjects. Each subject participated in a 2-class motor imagery EEG experiment, including left-hand (L) and right-hand (R) imagery. In this study, data from the first three sessions serve as the training set, while data from the remaining sessions served as the test set. Each subject completed 400 trials in the training set and 320 trials in the test set, resulting in a total of 720 trials per subject. The time segment for each trial is set to $\left[ 2.5,7\right] $ seconds.

\begin{table*}[!ht]
	\centering
	\caption{Classification Performance ($P_a (\%)$ and $K$) on BCI Competition \uppercase\expandafter{\romannumeral4}-2a for the Proposed EEG-DBNet with Other Models}
	\label{tab1}
	\setlength{\tabcolsep}{4pt}
	\begin{threeparttable}
		\begin{tabularx}{\textwidth}{Xc*{13}{r}}
			\toprule
			\multicolumn{1}{c}{\multirow{2}{*}{Subject}} & \multicolumn{2}{c}{EEGNet\cite{lawhern2018eegnet}} &
			\multicolumn{2}{c}{EEG-NeX\cite{chen2024toward}} &
			\multicolumn{2}{c}{EEG-ITNet\cite{salami2022eeg}} & \multicolumn{2}{c}{EEG-TCNet\cite{ingolfsson2020eeg}} &
			\multicolumn{2}{c}{MBEEGSE\cite{altuwaijri2022multi}} & 
			\multicolumn{2}{c}{ATCNet\cite{altaheri2022physics}} &  
			\multicolumn{2}{c}{EEG-DBNet}\\
			\cmidrule{2-15}
			& \multicolumn{1}{c}{$P_a$} & \multicolumn{1}{c}{$K$} & \multicolumn{1}{c}{$P_a$} & \multicolumn{1}{c}{$K$} & \multicolumn{1}{c}{$P_a$} & \multicolumn{1}{c}{$K$} & \multicolumn{1}{c}{$P_a$} & \multicolumn{1}{c}{$K$} & \multicolumn{1}{c}{$P_a$} & \multicolumn{1}{c}{$K$} & \multicolumn{1}{c}{$P_a$} & \multicolumn{1}{c}{$K$} & \multicolumn{1}{c}{$P_a$} & \multicolumn{1}{c}{$K$}\\
			\midrule
			\multicolumn{1}{c}{A1} & 89.32 & 0.8577 & 85.05 & 0.8007 & 86.48 & 0.8197 & 85.05 & 0.8007 & 90.39 & 0.8719 & 88.61 & 0.8481 & $\mathbf{90.75}$ & $\mathbf{0.8766}$ \\
			\multicolumn{1}{c}{A2} & 63.25 & 0.5103 & 71.02 & 0.6136 & 67.84 & 0.5714 & 64.31 & 0.5246 & 65.37 & 0.5386 & 73.85 & 0.6516 & $\mathbf{74.56}$ & $\mathbf{0.6610}$ \\
			\multicolumn{1}{c}{A3} & 94.87 & 0.9316 & 96.34 & 0.9512 & 95.60 & 0.9414 & 93.41 & 0.9121 & 95.60 & 0.9414 & 97.44 & 0.9658 & $\mathbf{97.44}$ & $\mathbf{0.9658}$ \\
			\multicolumn{1}{c}{A4} & 67.54 & 0.5666 & $\mathbf{85.09}$ & $\mathbf{0.8009}$ & 74.12 & 0.6543 & 70.18 & 0.6021 & 75.88 & 0.6781 & 81.58 & 0.7539 & 82.46 & 0.7659 \\
			\multicolumn{1}{c}{A5} & 75.00 & 0.6664 & 81.52 & 0.7535 & 77.90 & 0.7050 & 77.90 & 0.7053 & 77.54 & 0.6999 & 80.80 & 0.7443 & $\mathbf{82.61}$ & $\mathbf{0.7678}$ \\
			\multicolumn{1}{c}{A6} & 60.93 & 0.4792 & 63.72 & 0.5163 & 64.19 & 0.5229 & 62.79 & 0.5036 & 61.40 & 0.4852 & 73.95 & 0.6525 & $\mathbf{75.35}$ & $\mathbf{0.6713}$ \\
			\multicolumn{1}{c}{A7} & 89.89 & 0.8654 & 90.61 & 0.8749 & 90.97 & 0.8797 & 88.45 & 0.8461 & 93.14 & 0.9086 & $\mathbf{93.50}$ & $\mathbf{0.9134}$ & 91.34 & 0.8844 \\
			\multicolumn{1}{c}{A8} & 87.82 & 0.8377 & 83.39 & 0.7786 & 83.03 & 0.7737 & 82.66 & 0.7688 & 88.56 & 0.8475 & $\mathbf{90.04}$ & $\mathbf{0.8671}$ & 88.93 & 0.8524 \\
			\multicolumn{1}{c}{A9} & 81.44 & 0.7523 & 84.09 & 0.7878 & 85.23 & 0.8030 & 84.09 & 0.7878 & 85.98 & 0.8130 & $\mathbf{89.77}$ & $\mathbf{0.8635}$ & 89.39 & 0.8586 \\
			\midrule
			\multicolumn{1}{c}{St.D.\tnote{*}} & 11.93 & 0.1592 & 9.18 & 0.1224 & 9.86 & 0.1315 & 10.16 & 0.1355 & 11.52 & 0.1537 & 7.90 & 0.1053 & $\mathbf{7.23}$ & $\mathbf{0.0963}$ \\
			\multicolumn{1}{c}{Mean} & 78.90 & 0.7186 & 82.32 & 0.7642 & 80.60 & 0.7412 & 78.76 & 0.7168 & 81.54 & 0.7538 & 85.50 & 0.8067 & $\mathbf{85.87}$ & $\mathbf{0.8115}$ \\
			\bottomrule
		\end{tabularx}
		\begin{tablenotes}
			\footnotesize
			\item[*] Standard deviation.			
		\end{tablenotes}
	\end{threeparttable}
\end{table*}

\begin{table*}[!ht]
	\centering
	\caption{Classification Performance ($P_a (\%)$ and $K$) on BCI Competition \uppercase\expandafter{\romannumeral4}-2b for the Proposed EEG-DBNet with Other Models}
	\label{tab2}
	\setlength{\tabcolsep}{3.5pt} 
	\begin{threeparttable}
		\begin{tabularx}{\textwidth}{Xc*{13}{r}}
			\toprule
			\multicolumn{1}{c}{\multirow{2}{*}{Subject}} & \multicolumn{2}{c}{EEGNet\cite{lawhern2018eegnet}} &
			\multicolumn{2}{c}{EEG-NeX\cite{chen2024toward}} &
			\multicolumn{2}{c}{EEG-ITNet\cite{salami2022eeg}} & \multicolumn{2}{c}{EEG-TCNet\cite{ingolfsson2020eeg}} &
			\multicolumn{2}{c}{MBEEGSE\cite{altuwaijri2022multi}} & 
			\multicolumn{2}{c}{ATCNet\cite{altaheri2022physics}} &  
			\multicolumn{2}{c}{EEG-DBNet}\\
			\cmidrule{2-15}
			& \multicolumn{1}{c}{$P_a$} & \multicolumn{1}{c}{$K$} & \multicolumn{1}{c}{$P_a$} & \multicolumn{1}{c}{$K$} & \multicolumn{1}{c}{$P_a$} & \multicolumn{1}{c}{$K$} & \multicolumn{1}{c}{$P_a$} & \multicolumn{1}{c}{$K$} & \multicolumn{1}{c}{$P_a$} & \multicolumn{1}{c}{$K$} & \multicolumn{1}{c}{$P_a$} & \multicolumn{1}{c}{$K$} & \multicolumn{1}{c}{$P_a$} & \multicolumn{1}{c}{$K$}\\
			\midrule
			\multicolumn{1}{c}{B1} & 79.82 & 0.5970 & 80.70 & 0.6127 & 82.46 & 0.6498 & 80.26 & 0.6074 & 82.02 & 0.6420 & 81.58 & 0.6320 & $\mathbf{85.09}$ & $\mathbf{0.7019}$ \\
			\multicolumn{1}{c}{B2} & 71.43 & 0.4282 & $\mathbf{77.14}$ & $\mathbf{0.5428}$ & 75.51 & 0.5102 & 75.51 & 0.5101 & 71.43 & 0.4282 & 73.06 & 0.4607 & 75.10 & 0.5018 \\
			\multicolumn{1}{c}{B3} & 90.00 & 0.7998 & $\mathbf{92.61}$ & $\mathbf{0.8515}$ & 90.00 & 0.7979 & 90.87 & 0.8168 & 88.26 & 0.7652 & 90.87 & 0.8172 & 91.74 & 0.8339 \\
			\multicolumn{1}{c}{B4} & 98.37 & 0.9674 & $\mathbf{99.02}$ & $\mathbf{0.9805}$ & 98.70 & 0.9739 & 98.70 & 0.9739 & 98.70 & 0.9739 & 99.02 & 0.9805 & 98.37 & 0.9674 \\
			\multicolumn{1}{c}{B5} & 99.63 & 0.9927 & 99.63 & 0.9927 & 98.53 & 0.9707 & 98.53 & 0.9707 & 98.53 & 0.9707 & $\mathbf{100.00}$ & $\mathbf{1.0000}$ & 99.27 & 0.9853 \\
			\multicolumn{1}{c}{B6} & $\mathbf{93.63}$ & $\mathbf{0.8723}$ & 92.03 & 0.8406 & 90.44 & 0.8083 & 89.24 & 0.7842 & 92.03 & 0.8403 & 90.04 & 0.8009 & 92.43 & 0.8483 \\
			\multicolumn{1}{c}{B7} & $\mathbf{95.69}$ & $\mathbf{0.9138}$ & 94.83 & 0.8966 & 94.40 & 0.8879 & 94.83 & 0.8966 & 94.83 & 0.8966 & 96.55 & 0.9310 & 95.26 & 0.9052 \\
			\multicolumn{1}{c}{B8} & 95.22 & 0.9043 & 96.09 & 0.9217 & 96.52 & 0.9303 & 94.35 & 0.8869 & 95.22 & 0.9042 & 95.22 & 0.9043 & $\mathbf{96.96}$ & $\mathbf{0.9390}$ \\
			\multicolumn{1}{c}{B9} & 89.80 & 0.7959 & 89.80 & 0.7958 & 88.16 & 0.7629 & 89.80 & 0.7958 & $\mathbf{91.02}$ & $\mathbf{0.8203}$ & 90.20 & 0.8039 & 90.20 & 0.8038 \\
			\midrule
			\multicolumn{1}{c}{St.D.\tnote{*}} & 9.22 & 0.0721 & $\mathbf{7.12}$ & $\mathbf{0.0556}$ & 8.04 & 0.0630 & 8.14 & 0.0624 & 9.46 & 0.0805 & 8.92 & 0.0670 & 8.35 & 0.0661 \\
			\multicolumn{1}{c}{Mean} & 90.40 & 0.8079 & 91.32 & 0.8261 & 90.52 & 0.8102 & 90.23 & 0.8047 & 90.23 & 0.8046 & 90.73 & 0.8145 & $\mathbf{91.60}$ & $\mathbf{0.8318}$ \\			
			\bottomrule
		\end{tabularx}
		\begin{tablenotes}
			\footnotesize
			\item[*] Standard deviation.			
		\end{tablenotes}
	\end{threeparttable}
\end{table*}

\subsection{Experiment Settings}
\label{sec: Experiment Settings}

The proposed EEG-DBNet and other comparative models were trained and tested using the TensorFlow framework, running on the Intel(R) Xeon(R) Gold 6248R CPU @ 3.00GHz and NVIDIA A10 GPU. Both datasets utilized data from all EEG channels, excluding three electrooculography (EOG) channels. All models were trained using the Adam optimizer with a learning rate of 0.0009, a batch size of 64, and cross-entropy loss over 1000 epochs with early stopping set to patience of 300 epochs. The best-performing round out of ten training rounds, based on test results, was selected for evaluation.

The evaluation metrics used were classification accuracy $P_a$ and the kappa value $K$. The classification accuracy $P_a$ is defined as:
\begin{equation}
	\label{eq19}
	P_a=\frac{\sum_{n}^{i=1}\text{TP}_i/l_i}{n}
\end{equation}
where $\text{TP}_i$ denotes the number of correctly predicted (true positive) samples in class $i$, $l_i$ is the number of samples in class $i$, and $n$ is the number of classifications. The kappa value $K$ is defined as:
\begin{equation}
	\label{eq20}
	K=\frac{P_a-P}{1-P}
\end{equation}
where $P$ indicates the hypothetical probability of chance agreement.

\subsection{Performance Comparison}
\label{sec: Performance Comparison}

The proposed EEG-DBNet is compared with six state-of-the-art methods in recent years: EEGNet\cite{lawhern2018eegnet}, EEG-NeX\cite{chen2024toward}, EEG-ITNet\cite{salami2022eeg}, EEG-TCNet\cite{ingolfsson2020eeg}, MBEEGSE\cite{altuwaijri2022multi}, and ATCNet\cite{altaheri2022physics}. EEGNet, a classic method for decoding MI-EEG, is widely regarded as a benchmark. EEG-NeX is an improved version of EEGNet. The other models share structural similarities with the proposed method. EEG-ITNet and MBEEGSE utilize a branched architecture, while EEG-TCNet and ATCNet include modules for extracting global features. We reproduced all these methods on the BCI Competition \uppercase\expandafter{\romannumeral4}-2a and \uppercase\expandafter{\romannumeral4}-2b datasets based on the model parameters described in the original papers.

\begin{figure}[!h]
	\begin{center}
		\subfigure[]{\includegraphics{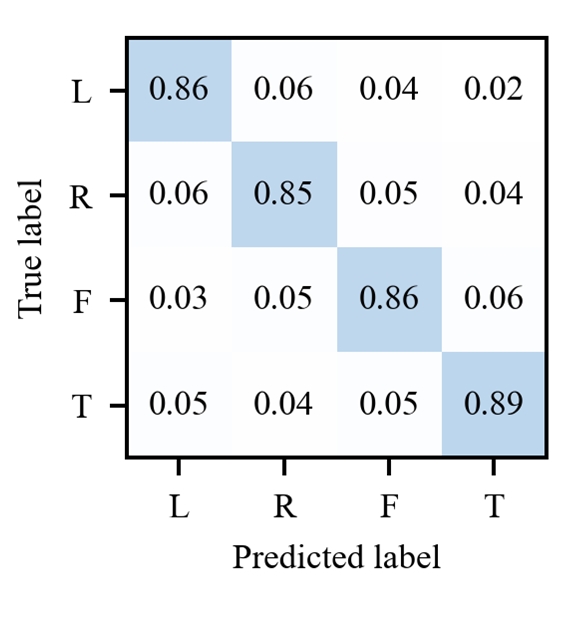}\label{fig_4a}}
		\hfill
		\subfigure[]{\includegraphics{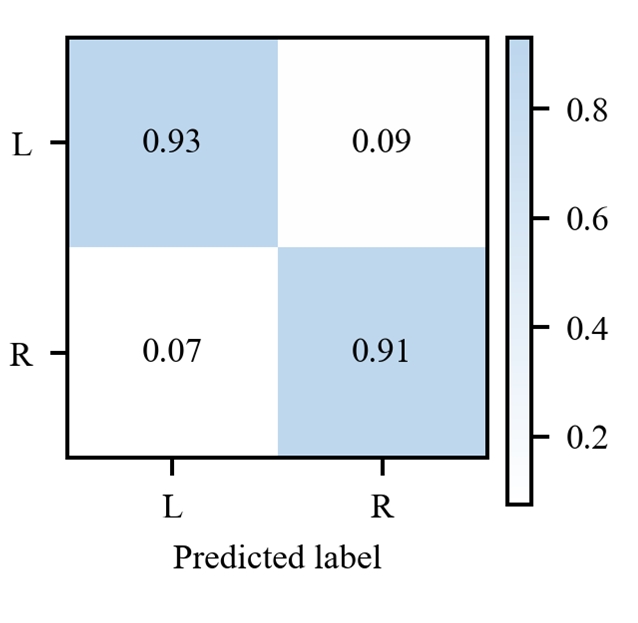}\label{fig_4b}}
		\caption{Confusion matrices of EEG-DBNet. ``L'' stands for left hand, ``R'' stands for right hand, ``F'' stands for feet, and ``T'' stands for tongue. \subref{fig_4a} BCI Competition \uppercase\expandafter{\romannumeral4}-2a. \subref{fig_4b} BCI Competition \uppercase\expandafter{\romannumeral4}-2b.}
		\label{fig_4}
	\end{center}
\end{figure}

\autoref{tab1} and \autoref{tab2} present the experimental results for each model on the BCI Competition \uppercase\expandafter{\romannumeral4}-2a and \uppercase\expandafter{\romannumeral4}-2b datasets, respectively. EEG-DBNet achieved the best results on the BCI Competition \uppercase\expandafter{\romannumeral4}-2a dataset. ATCNet's performance was close to that of EEG-DBNet, with a significant increase in accuracy compared to other methods. Notably, EEG-DBNet also had the lowest standard deviation, indicating its superior generalization capability. Similarly, EEG-DBNet achieved the best results on the BCI Competition \uppercase\expandafter{\romannumeral4}-2b dataset. Given that the BCI Competition \uppercase\expandafter{\romannumeral4}-2b dataset is derived from a binary classification task, which is simpler compared to the four-class classification in BCI Competition \uppercase\expandafter{\romannumeral4}-2a, all methods attained relatively high classification accuracy. In this dataset, EEG-NeX performed comparably to EEG-DBNet and demonstrated the best generalization. EEG-DBNet outperformed other methods by a margin of one percentage point. Overall, the proposed EEG-DBNet achieved optimal results on both datasets, demonstrating its excellent EEG decoding capabilities. The confusion matrix for EEG-DBNet is shown in \autoref{fig_4}.

\subsection{Ablation Experiments}
\label{sec: Ablation Experiments}

The effectiveness of each module of EEG-DBNet was validated through ablation experiments on the BCI Competition \uppercase\expandafter{\romannumeral4}-2a. As shown in \autoref{tab3}, the dual-branch network, which employs average pooling for temporal sequences and max pooling for spectral sequences (LC block), improves accuracy by $3.5\%$ over the single-branch network, achieving $82.43\%$. This indicates that the dual-branch network has superior feature extraction capabilities, which are achieved by constructing diversified CNNs. The introduction of DCCNNs further enhances accuracy by $0.35\%$, reaching $82.78\%$, demonstrating that DCCNNs serve as a valuable supplement to CNNs. The key difference between DCCNNs and CNNs is that DCCNNs increase the receptive field by adding network depth, thereby enhancing the network's global feature extraction ability. 

\begin{table}[!h]
	\centering
	\caption{Contribution of EEG-DBNet blocks for BCI Competition \uppercase\expandafter{\romannumeral4}-2a}
	\label{tab3}
	\begin{threeparttable}
		\begin{tabularx}{\columnwidth}{Xrr}
			\toprule
			\multicolumn{1}{c}{Block} & $\bar{P_a}(\%)$\tnote{*} & $\bar{K}$\tnote{*} \\
			\midrule
			Single-branch (EEGNet\cite{lawhern2018eegnet})& 78.90 & 0.7186\\
			Dual-branch (LC\tnote{1}) & 82.43 & 0.7656\\			
			LC\tnote{1}+GC without SE and SW & 82.78 & 0.7703\\
			LC\tnote{1}+GC without SE & 84.39 & 0.7918\\
			LC\tnote{1}+GC without SW & 83.61 & 0.7814\\
			LC\tnote{2}+GC & 83.16 & 0.7754\\
			LC\tnote{3}+GC & 82.85 & 0.7713\\
			LC\tnote{4}+GC & 81.90 & 0.7587\\
			LC\tnote{1}+GC (EEG-DBNet)  & $\mathbf{85.87}$ & $\mathbf{0.8115}$\\
			\bottomrule
		\end{tabularx}
		\begin{tablenotes}
			\footnotesize
			\item[*] Average of nine subjects.
			\item[1] Average pooling for temporal branch and max pooling for spectral branch.
			\item[2] Average pooling for both branches.
			\item[3] Max pooling for both branches.
			\item[4] Max pooling for temporal branch and average pooling for spectral branch.
		\end{tablenotes}
	\end{threeparttable}
\end{table}

Splitting the sequence into six subsequences and then extracting features from these using DCCNNs improves the model's classification accuracy by $1.61\%$, reaching $84.39\%$. This suggests that sequence splitting enables DCCNNs to construct different feature extraction paradigms and increase feature diversity. By reconstructing the features of the sequence and then extracting features using DCCNNs, the model's classification accuracy improves by $0.83\%$, reaching $83.61\%$. This indicates that remodeling the relationships between sequence samples helps DCCNNs more accurately extract signal features. When both sequence splitting and feature reconstruction are used simultaneously, DCCNNs processing the reconstructed subsequences can improve the model's classification accuracy by $3.09\%$, surpassing the sum of their individual effects ($1.61\% + 0.83\% = 2.44\%$), reaching $85.87\%$. This demonstrates that reconstructed subsequences can enhance the efficiency of DCCNNs in constructing different feature extraction paradigms. The introduction of the GC block based on DCCNNs significantly improves model performance, increasing accuracy by $3.44\%$ over the dual-branch network with only the LC block, indicating that the GC block and LC block provide substantial complementary effects.

Furthermore, \autoref{tab3} shows that the model performance is highly sensitive to the type of pooling layer used for different signal dimensions. If both network branches utilize average pooling, the accuracy of the model decreases by $2.71\%$ compared to EEG-DBNet, reaching $83.16\%$. If both network branches use max pooling, the accuracy decreases by $3.02\%$, dropping to $82.85\%$. When max pooling is used in the temporal branch and average pooling in the spectral branch, accuracy drops by $3.97\%$, reaching $81.90\%$, which is even lower than the performance of the network composed solely of LC. Therefore, we deduce that for spectral sequences, specific sample points better reflect the sequence characteristics than the relationships between multiple sample points. Conversely, for temporal sequences, the relationships between sample points are more important than individual points. As a result, average pooling is more suitable for temporal sequences, while max pooling is more appropriate for spectral sequences.

\subsection{Parameter Sensitivity}
\label{sec: Parameter Sensitivity}

\begin{figure}[!h]
	\centering
	\includegraphics{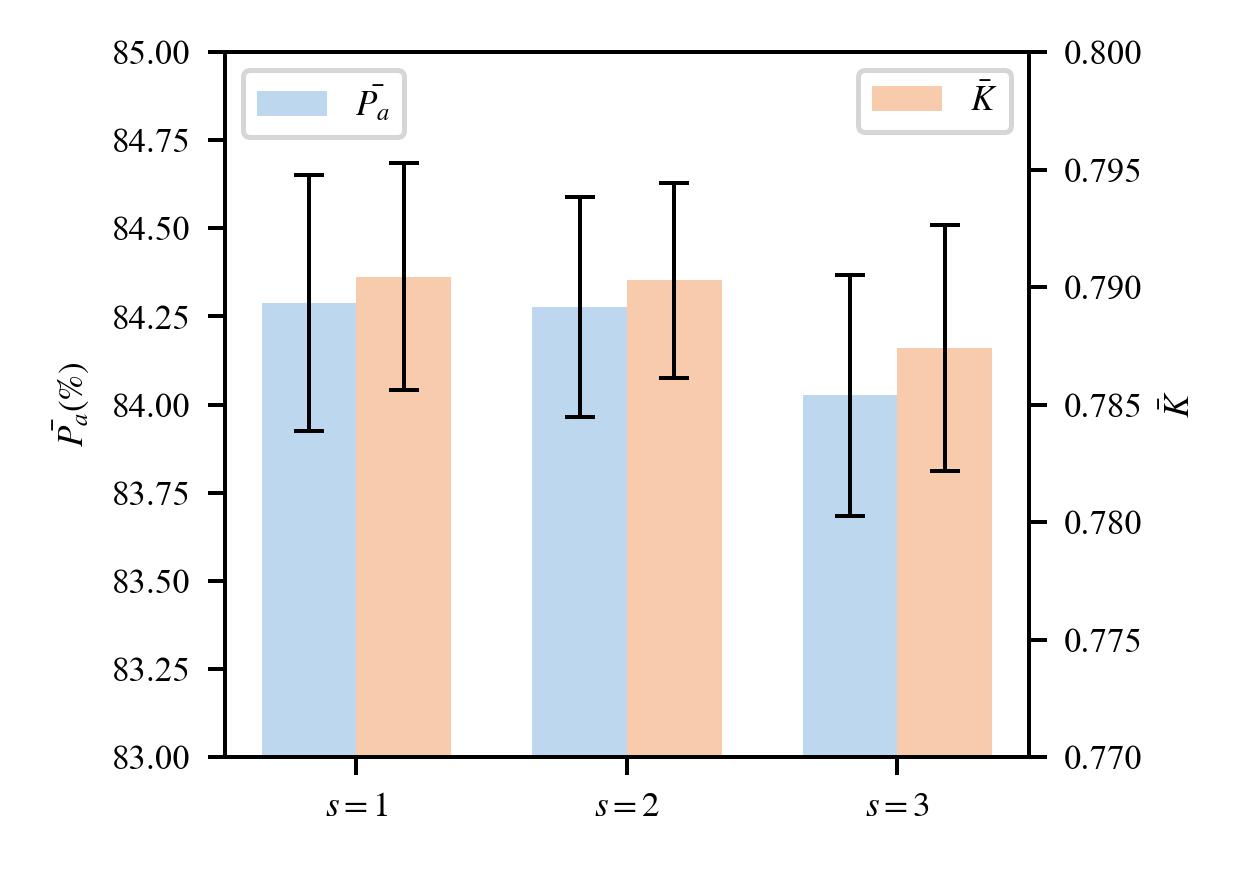}
	\caption{Classification performance ($P_a (\%)$ and $K$) on BCI Competition \uppercase\expandafter{\romannumeral4}-2a for different values of $s$.}
	\label{fig_5}
\end{figure}

\begin{figure}[!h]
	\centering
	\includegraphics{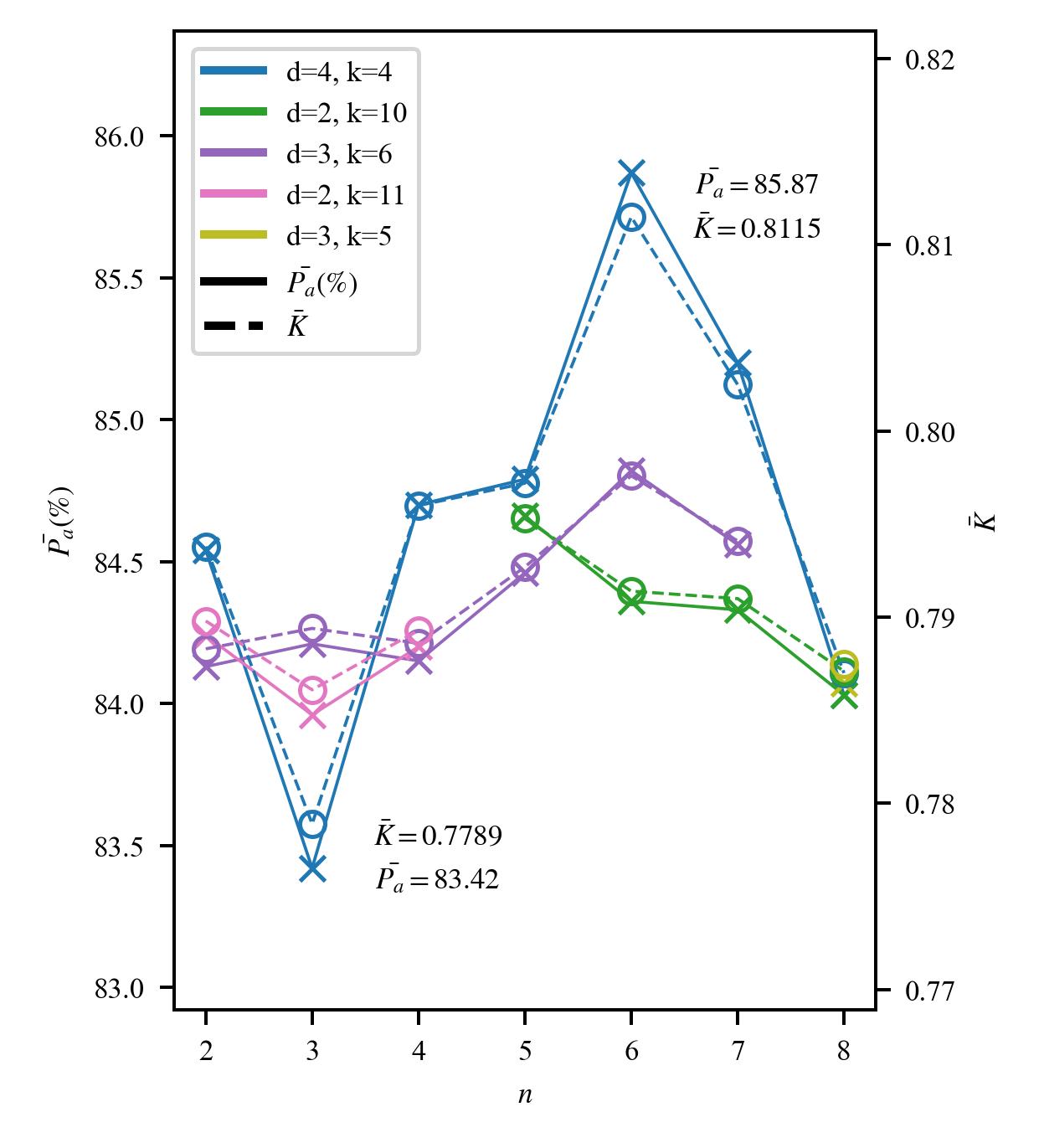}
	\caption{Classification performance ($P_a (\%)$ and $K$) on BCI Competition \uppercase\expandafter{\romannumeral4}-2a for various hyperparameters ($s=1$).}
	\label{fig_6}
\end{figure}

The GC block has four hyperparameters: the stride $s$ of the sliding window, the number of sliding windows $n$, the kernel size $k$ of the dilated convolution, and the number of dilated convolution layers $d$. The first step of the GC block is to split the sequence, where $s$ must be determined to segment the sequence into $n$ parts based on its length. 
We conducted 36 experiments on the BCI Competition \uppercase\expandafter{\romannumeral4}-2a with $s\in\{1,2,3\}$ and $n\in\{2,3,4,5\}$, matching the conresponding $k$ and $d$ according to the constraints of (\ref{eq16}). We recorded the average $P_a$ ($\bar{P_a}$) and average $K$ ($\bar{K}$) across nine subjects to evaluate the impact of different $s$ values on the GC block. The experimental results are illustrated in \autoref{fig_5}, where the results of the 36 experiments are divided into three groups based on different $s$ values. It can be observed that as the $s$ increases, the classification accuracy of the model on the signal decreases. This is likely due to the larger stride accentuating the differences in the features of each subsequence. Therefore, we set $s=1$, and then choose $n\in\{2,3,4,5,6,7,8\}$ to select all $d$ and $k$ that satisfy the constrains of (\ref{eq16}) for our experiments. The results, as shown in \autoref{fig_6}, indicate that $d=4$ and $k=4$ meet the constraint conditions for all values of $n$, while $d=3$ and $k=5$ only satisfy the constraints when $n=8$. The model exhibits its worst and best accuracy both at $d=4$ and $k=4$. Specifically, the model performs worst at $d=4$, $k=4$, $n=3$ with $\bar{P_a}=83.42\%$ and $\bar{K}=0.7789$, and it performs best at $d=4$, $k=4$, $n=6$ with $\bar{P_a}=85.87\%$ and $\bar{K}=0.8115$. Consequently, we set the four hyperparameters of the GC block to $s=1$, $d=4$, $k=4$, and $n=6$.

\section{Conclusion and Discussion}
\label{sec: Conclusion and Discussion}

This study proposes EEG-DBNet, a dual-branch network structure designed to parallelly decode the temporal and spectral sequences of MI-EEG, accurately discerning subjects' motor intentions. The temporal sequences alone provide limited signal features; thus, a multi-dimensional exploration of these features is crucial for enhancing model performance. The parallel dual-branch network is essential for this decoding approach. If temporal and spectral sequences are decoded sequentially, the model accuracy drops below $81\%$, highlighting the equal importance of both feature types for decoding MI-EEG.

The primary function of the LC block is to remodel sequence sample points. During this process, the LC block captures local signal features and adjusts sequence length by decoding inter-sample relationships, setting the stage for subsequent global feature extraction. Our findings indicate that using different types of pooling layers for sequences of varying dimensions significantly impacts model performance. Specifically, employing an average pooling layer in the temporal branch and a max pooling layer in the frequency branch enables optimal model performance. We infer that incorporating information from all samples within the pooling range better represents the temporal sequence characteristics, while the maximum value of the samples within the pooling range more effectively captures signal fluctuation amplitudes, thereby better representing frequency sequence characteristics.

The value of DCCNNs lies in their ability to adjust the network's receptive field, addressing the ``breadth'' issue in signal decoding. Sequence splitting allows DCCNNs to train multiple signal decoding models, extracting features from each subsequence and thus enhancing the network's ability to capture global features in terms of ``depth.'' High similarity between subsequences hinders the training of diverse feature extraction models in DCCNNs. Feature reconstruction of subsequences increases their diversity while maintaining continuity, thereby improving the model's efficiency in extracting global features. Sequence splitting and feature reconstruction effectively strengthen DCCNNs, and fine-tuning DCCNN hyperparameters according to sequence length significantly improves model performance.

In summary, the design of EEG-DBNet is effective, with each module significantly enhancing model performance. Experiments on two public datasets achieved optimal results. However, the numerous hyperparameters in EEG-DBNet make the tuning process cumbersome. Integrating the LC and GC blocks to reduce hyperparameters could improve the model's practicality. Additionally, exploring methods for extracting spatial signal features to further increase signal decoding dimensions is a key direction for future research.


\bibliographystyle{elsarticle-num-names} 
\bibliography{bibliographies}


%
%
%
\end{document}